\documentclass[openacc]{rstransa}

\begin{document}

\title{The Evolution of Massive Stars: Bridging the Gap in the Local Group}

\author{
Philip Massey$^{1,2}$, Kathryn F. Neugent$^{1,2}$, and Emily M. Levesque$^{3}$}

\address{$^{1}$Lowell Observatory, 1400 W Mars Hill Road, Flagstaff, AZ 86001, USA\\
$^2$Department of Physics and Astronomy, Northern Arizona University, Box 6010, Flagstaff, AZ 86011-6010, USA\\
$^3$Department of Astronomy, University of Washington, Box 351580, Seattle, WA 98195, USA\\}

\subject{astrophysics}


\corres{Philip Massey\\
\email{phil.massey@lowell.edu}}

\begin{abstract}
The nearby galaxies of the Local Group  can act as our laboratories in helping to bridge
the gap between theory and observations.  In this review we will describe the complications of identifying samples of
OB stars, yellow and red supergiants, and Wolf-Rayet stars, and  what we have so far learned from these studies. 
\end{abstract}


\begin{fmtext}
\section{Introduction}
In calling this conference ``Bridging the Gap," the organizers perhaps had in mind the gap in our knowledge between supernovae and their progenitors, but our focus here will be on the gap between theory and observations, and bridging it
using observations of stars in the nearby galaxies of the Local Group.  These galaxies can serve as our natural astrophysical laboratories, as massive star evolution depends strongly on mass loss, and the mass-loss rates on the main-sequence depend heavily on metallicity (see, e.g., [1]). The star-forming galaxies of the Local Group span a range of 25 in metallicity, from the metal poor Sextans A and B galaxies, with a metallicity about 0.06$\times$ solar [2],  to the metal-rich Andromeda Galaxy, M31, with a metallicity that is approximately 1.6$\times$ solar [3,4].

\end{fmtext}

\maketitle

This review will focus on explaining how we find massive stars in nearby galaxies, the inventory and completeness of the samples we have found, and what we have learned by finding them. We have  organized it as follows: Section 2,  O-type stars and B supergiants; Section 3, yellow and red supergiants; and Section 4, Wolf-Rayet (WR) stars.  The Luminous Blue Variable (LBV) stars will be discussed elsewhere in this volume by Nathan Smith.

\section{OB Stars}

One of the nice things about looking for OB stars in nearby galaxies is that there is virtually no foreground contamination: one can be pretty sure that
blue stars in the right magnitude range really are members of that galaxy.  There are, however, two down sides.  First, 
these stars are so hot that their observable light is on the tail of the Rayleigh-Jeans distribution, and there is very little difference in the colors of these
stars over a significant temperature range (30,000-50,000 K).  Yet the bolometric corrections are quite sensitive to effective temperature, and so converting the observables to physical properties reliably requires spectroscopy once a suitable sample has been identified [5].

The second problem is that these stars are fainter than their evolved
descendants, the yellow and red supergiants, owing to the fact that massive stars evolve at fairly constant bolometric luminosities and that OB stars are so
hot that most of their flux is in the far-ultraviolet.  This is particularly true for the most interesting of these massive progenitors, the young (zero-age) massive O-type
stars.

We illustrate this in Table~\ref{table:evol} where we use 
solar-metallicity single-star evolutionary tracks of the Geneva group [6] to list the expected stellar parameters as a function of age.  This is based upon Table 1 of [5], but uses models that are 20 years more advanced.

\begin{figure}[!h]
\centering\includegraphics[width=2.2in]{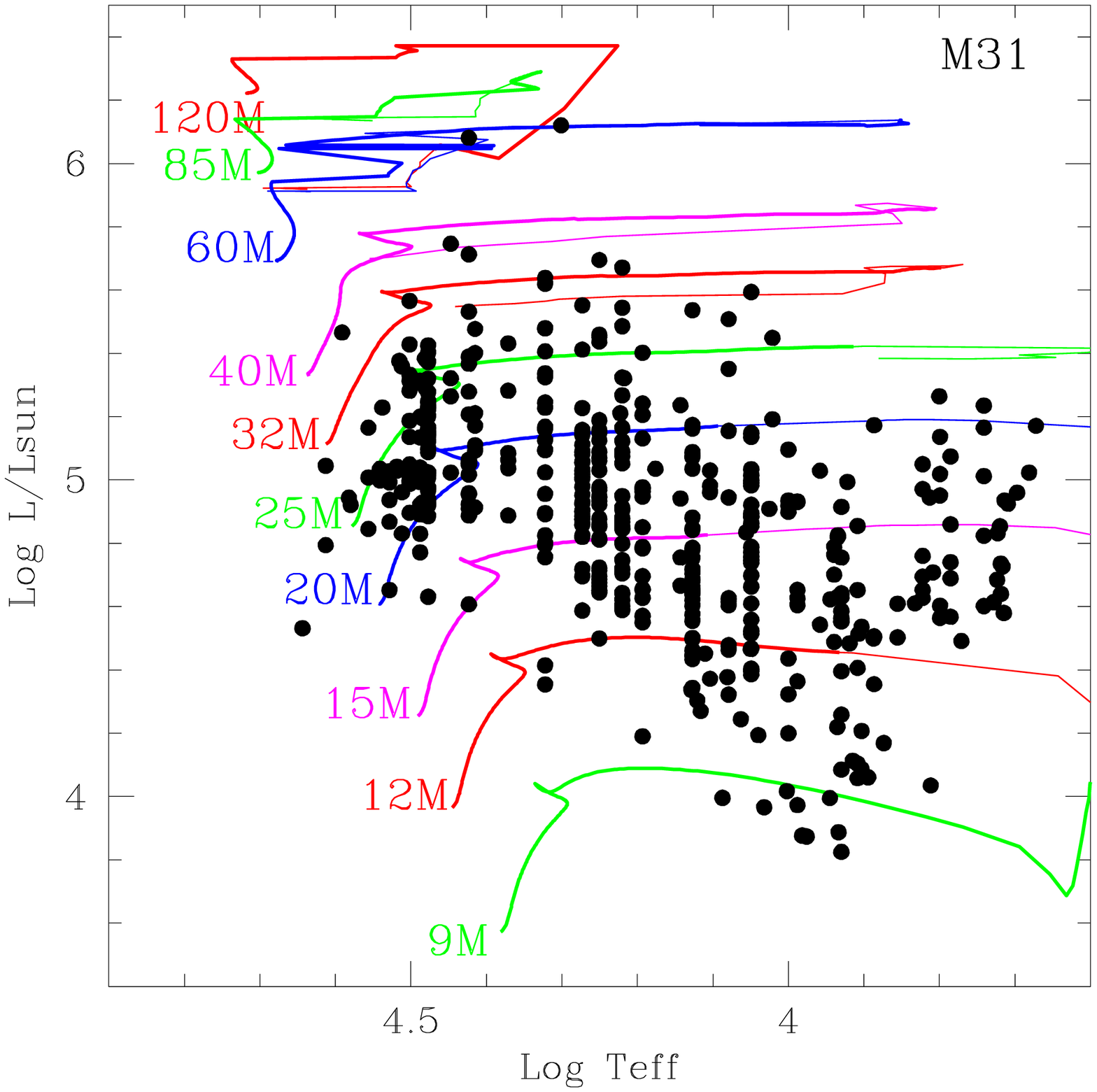}
\centering\includegraphics[width=2.2in]{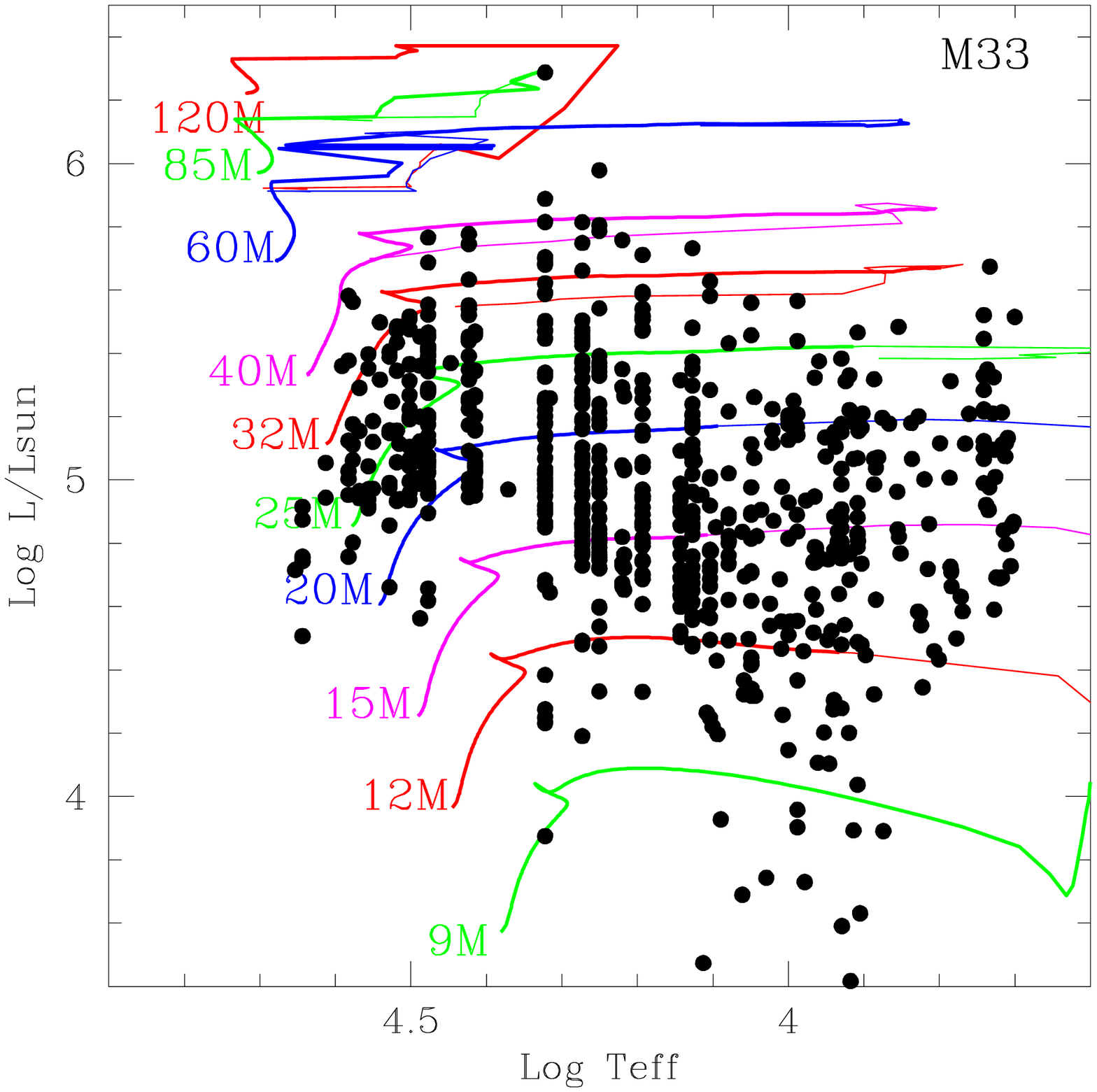}
\caption{The H-R diagrams for M31 and M33 showing only stars with known spectral types. Note the lack of spectroscopically observed high mass stars, and the overall scant number of O-type stars ($\log T_{\rm eff}>4.48$). From [7]; used
with permission.}
\label{fig:HRDs}
\end{figure}

Massive stars evolve at (more or less) constant $\log L/L_\odot$. As the star evolves, the star expands, the effective temperature cools, and the surface gravity decreases.  The star becomes visually brighter because the peak of the spectral energy distribution shifts to longer wavelengths as the temperature cools.  Thus a 6~Myr old 25$M_\odot$ star (spectroscopically, an O8.5~I) has an absolute visual magnitude $M_V=-5.0$.  This late-middle-age 25$M_\odot$ star will thus be just as bright visually as an zero-age main-sequence 60$M_\odot$ (O3~V) star!  Since 25$M_\odot$ stars are significantly more common than those of 60$M_\odot$, we expect these stars to dominate in any magnitude-limited sample.
Indeed, this fact is cited by [7] as the reason that the highest mass stars are conspicuously absent in the H-R diagram shown in Fig.~\ref{fig:HRDs}

\begin{table}[!h]
\scriptsize
\caption{Evolution of Massive Stars at Solar Metallicity According to [6]}
\label{table:evol}
\begin{tabular}{l c c c c c c c c}
\multicolumn{9}{c}{120$M_\odot$} \\
\multicolumn{9}{c}{$\tau_{\rm ms}$=3.1 Myr} \\
\multicolumn{9}{c}{Age (Myr)} \\ \cline{2-9}
&  0.0 &   0.5 &   1.0 &   1.5 &   2.0 &   2.5 &   3.0 &   3.5 \\ \hline
Log T$_{\rm eff}$ & 4.715& 4.704& 4.713& 4.726& 4.736& 4.514& 4.520& 4.745\\
Log $L/L_\odot$ & 6.22& 6.25& 6.27& 6.30& 6.32& 6.34& 6.37& 5.80\\
Log g [cgs] &  4.1&  4.0&  4.0&  4.0&  4.0&  3.0&  2.9&  3.9\\
$M_V$ & -6.3& -6.4& -6.4& -6.4& -6.4& -7.9& -8.0& -5.0\\
Sp Type &O2-3~V & O2-3~V & O2-3~V & O2-3~V & O2-3~V & WR & WR & WR \\
\hline
\end{tabular}
\vskip 10pt

\begin{tabular}{l c c c c c c c c c}
\multicolumn{10}{c}{85$M_\odot$} \\
\multicolumn{10}{c}{$\tau_{\rm ms}$=3.7 Myr} \\
\multicolumn{10}{c}{Age (Myr)} \\ \cline{2-10}
&  0.0 &   0.5 &   1.0 &   1.5 &   2.0 &   2.5 &   3.0 &   3.5 &   4.0 \\ \hline
Log T$_{\rm eff}$ & 4.700& 4.687& 4.684& 4.689& 4.704& 4.723& 4.546& 4.530& 4.728\\
Log $L/L_\odot$ & 5.97& 5.99& 6.02& 6.05& 6.08& 6.12& 6.15& 6.19& 6.01\\
Log g [cgs] &  4.2&  4.1&  4.0&  4.0&  4.0&  4.0&  3.2&  3.1&  3.7\\
$M_V$ & -5.7& -5.9& -6.0& -6.0& -6.0& -6.0& -7.3& -7.5& -5.6\\
Sp Type &O2-3~V & O2-3~V & O2-3~V & O2-3~V & O2-3~V & O2-3~V & WR & WR & WR \\
\hline
\end{tabular}
\vskip 10pt

\begin{tabular}{l c c c c c c c c c c}
\multicolumn{11}{c}{60$M_\odot$} \\
\multicolumn{11}{c}{$\tau_{\rm ms}$=4.5 Myr} \\
\multicolumn{11}{c}{Age (Myr)} \\ \cline{2-11}
&  0.0 &   0.5 &   1.0 &   1.5 &   2.0 &   2.5 &   3.0 &   3.5 &   4.0 &   4.5 \\
\hline
Log T$_{\rm eff}$ & 4.675& 4.666& 4.659& 4.655& 4.657& 4.664& 4.676& 4.684& 4.529& 4.165\\
Log $L/L_\odot$ & 5.69& 5.71& 5.74& 5.77& 5.81& 5.84& 5.88& 5.92& 5.97& 6.12\\
Log g [cgs] &  4.2&  4.1&  4.1&  4.0&  3.9&  3.9&  3.9&  3.9&  3.2&  1.5\\
$M_V$ & -5.2& -5.3& -5.4& -5.5& -5.6& -5.7& -5.7& -5.7& -6.9& -9.8\\
Sp Type &O3~V & O3~V & O3~V & O3~V & O3~V & O3~V & O3~V & O3~V & WR & WR \\
\hline
\end{tabular}
\vskip 10pt

\begin{tabular}{l c c c c c c c }
\multicolumn{8}{c}{40$M_\odot$} \\
\multicolumn{8}{c}{$\tau_{\rm ms}$=5.7 Myr} \\
\multicolumn{8}{c}{Age (Myr)} \\ \cline{2-8}
&  0.0 &   1.0 &   2.0 &   3.0 &   4.0 &   5.0 &   6.0  \\
\hline
Log T$_{\rm eff}$ & 4.635& 4.624& 4.613& 4.601& 4.594& 4.579& 4.618\\
Log $L/L_\odot$ & 5.33& 5.38& 5.44& 5.50& 5.58& 5.66& 5.61\\
Log g [cgs] &  4.2&  4.1&  4.0&  3.9&  3.8&  3.6&  3.5\\
$M_V$ & -4.6& -4.8& -5.0& -5.2& -5.5& -5.8& -5.4\\
Sp Type &O4~V& O5 V & O5 V&  O5.5 V &  O5.5 III & O5 I  & WR \\
\hline
\end{tabular}
\vskip 10pt

\begin{tabular}{l c c c c c c c c c c c }
\multicolumn{12}{c}{25$M_\odot$} \\
\multicolumn{12}{c}{$\tau_{\rm ms}$=8.0 Myr} \\
\multicolumn{12}{c}{Age (Myr)} \\ \cline{2-12}
&  0.0 &   1.0 &   2.0 &   3.0 &   4.0 &   5.0 &   6.0 &   7.0 &   8.0 & 8.1 & 8.5 \\
\hline
Log T$_{\rm eff}$ & 4.578& 4.569& 4.564& 4.557& 4.546& 4.533& 4.513& 4.480& 3.584 & 3.861 & 4.531\\
Log $L/L_\odot$ & 4.85& 4.89& 4.93& 4.98& 5.03& 5.09& 5.15& 5.23& 5.42 &5.38 & 5.35\\
Log g [cgs] &  4.3&  4.2&  4.1&  4.0&  3.9&  3.8&  3.7&  3.5& -0.4 & 0.6 & 3.2 \\
$M_V$ & -3.8& -3.9& -4.1& -4.2& -4.4& -4.7& -5.0& -5.4&-9.8 & -8.7 & -5.4\\
Sp Type &O6 V  &O6.5 V & O6.5 V & O7 V & O7.5 V &  O8 III & O8.5 I & B0 I & K5-M0~I & F2~I & WR \\
\hline
\end{tabular}
\vskip 10pt

\begin{tabular}{l c c c c c c c c}
\multicolumn{9}{c}{20$M_\odot$} \\
\multicolumn{9}{c}{$\tau_{\rm ms}$=9.6 Myr} \\
\multicolumn{9}{c}{Age (Myr)} \\ \cline{2-9}
  &0.0 &   2.0 &   4.0 &     6.0 &    8.0 &    9.8 &10.0&  10.2\\
\hline
Log T$_{\rm eff}$ & 4.542&  4.531& 4.522&  4.504&  4.466&  3.750 & 3.581 & 3.815\\
Log $L/L_\odot$ & 4.61&  4.66 4.74&  4.83&  4.94& 5.19& 5.15 & 5.13\\
Log g [cgs] &  4.3&   4.2&    4.0&  3.9&  3.6&   0.5 & -0.3 & 0.5 \\
$M_V$ & -3.4& -3.6&  -3.9&  -4.2&  -4.8&  -8.1 &-9.1 & -8.1\\
Sp Type &O7.5 V &  O8 V &O8.5 V & O9 V & B0 I & G0 I & M0~I & F5~I\\
\hline
\end{tabular}
\vskip 10pt

\begin{tabular}{l c c c c c c c c c }
\multicolumn{10}{c}{15$M_\odot$} \\
\multicolumn{10}{c}{$\tau_{\rm ms}$=13.6 Myr} \\
\multicolumn{10}{c}{Age (Myr)} \\ \cline{2-10}
&  0.0 &   2.0 &   4.0 & 6.0 & 8.0 & 10.0 &  12.0 &  14.0 &  15.0 \\
\hline
Log T$_{\rm eff}$ & 4.490&  4.482& 4.477&  4.471& 4.460&  4.444&  4.413& 3.571& 3.562\\
Log $L/L_\odot$ & 4.25& 4.29& 4.34& 4.39& 4.46& 4.54&  4.62& 4.82& 4.93 \\
Log g [cgs] &  4.3&   4.2&  4.1&  4.1&   3.9&  3.8&  3.6&   0.0& -0.2 \\
$M_V$ & -2.9&  -3.0& -3.2& -3.4& -3.6&  -3.9& -4.3& -8.1 & -8.2\\
Sp Type &O9.5 V & B0 V& B0 V& B0 V  & B0 III & B0 I & B0.2 I & M1.5 I &M2 I \\
\hline
\end{tabular}
\vskip 10pt

\begin{tabular}{l c c c c c c c c c c c }
\multicolumn{12}{c}{12$M_\odot$} \\
\multicolumn{12}{c}{$\tau_{\rm ms}$=18.5 Myr} \\
\multicolumn{12}{c}{Age (Myr)} \\ \cline{2-12}
&  0.0 &   2.0 &   4.0 &   6.0 &   8.0 &  10.0 &  12.0 &  14.0 &  16.0 &  18.0 &  20.0 \\ \hline
Log T$_{\rm eff}$ 4.445& & 4.438& 4.434& 4.431& 4.427& 4.421& 4.412 & 4.399 & 4.381 & 4.349 & 3.567\\
Log $L/L_\odot$ & 3.97 & 3.99 & 4.03 & 4.06 & 4.10 & 4.14 & 4.19 & 4.25 & 4.31 & 4.39 & 4.56 \\
Log g [cgs] &  4.3&  4.2&  4.2&  4.1&  4.1&  4.0&  3.9&  3.8&  3.7&  3.5&  0.1\\
$M_V$ & -2.5& -2.6& -2.7& -2.8& -2.9& -3.1& -3.3& -3.5& -3.8& -4.2&-7.4\\
Sp Type &B0.2 V & B0.2 V & B0.2 V & B0.2 V & B0.5 V  & B0.5 V& B0.2~I & B0.2 I & B0.5 I & B0.7 I & M1.5~I \\
\hline 
\end{tabular}
\end{table}

\clearpage

However, there is a far larger problem revealed by Fig.~\ref{fig:HRDs}.   We have so far identified only 64 O-type stars in M31, and 130 in M33 [7].   These numbers can be contrasted with those expected given the number
of WRs in these galaxies, 154 and 206, respectively [7].   (The fact that M31 is a much larger galaxy than M33 but has fewer WRs is a reflection of its lower star formation rate.) According to the Geneva single-star evolutionary models [8], there should be 15$\times$ more
O-type stars than WRs, suggesting that  there should be 2000-3000 O-type stars present.  Where are the others?  Since the problem occurs for both M31 and M33 we believe this is {\it not} telling us something profound about massive evolution.  It is the hotter, hydrogen-burning stars that are missing. Rather, these samples have been biased towards the later-type stars.  We hope to rectify this soon, using new data we are obtaining in the UV with {\it HST}.

\section{Yellow and Red Supergiants}

The {\it evolved} massive stars provide an extraordinary sensitive test of massive star evolutionary theory.  Their numbers and physical properties are sensitive to the details of earlier stages, and thus they can act as a ``magnifying glass" to illustrate any faults in the calculations of earlier stages in the evolutionary models [9].  This applies not only to yellow supergiants (YSGs) and red supergiants (RSGs), but also to the subject of the next section, the WRs.

However, a problem that applies to the YSGs and RSGs, that does not affect the OB stars or the WRs, is that issue of foreground contamination by stars with similar colors and
magnitudes.  The YSGs range from  $M_V=-6.5$ (12$M_\odot$) to $-9.5$ (25$M_\odot$); in the Magellanic Clouds, this translates to visual magnitudes of $V\sim 9.5-12.5.$
Foreground yellow dwarfs ($M_V\sim+4$) at a distance of 100-500 pc will thus fall in the appropriate magnitude range.  In the more distant galaxies of the Local Group,
such as M31 and M33, the YSGs will be found at $V\sim 15.5-18.5$, and the corresponding contaminating foreground stars would be found at distances of 2-8~kpc. The issue with
RSGs is a little more complicated, as the bolometric corrections depend upon the exact spectral types, but roughly at M0~I ($M_V=-5$ at 12M$_\odot$ and -8.0 at 30$M_\odot$), 
RSGs will be found at $V=11.0-14.0$ in the Magellanic Clouds, and $V=17-20$ in M31 and M33. A foreground M0~V has $M_V=+9$, and hence at distances of 25-100 pc (!) for the
Magellanic Clouds, and 400-1500 for M31 and M33.
Empirically, what we've found by detailed radial velocity studies is that
the contamination of the YSGs is so high in the Magellanic Clouds that even after eliminating stars with significant proper motions,  60-80\% of the sample proved to be foreground
stars [10,11], while the contamination is even higher ($>$90\%) in M31 [12] and M33 [13].  For the RSGs in the Magellanic Clouds, elimination of the sample via proper motions proved
quite effective (not surprising, given their close distances!) [11], and for the samples in M31 and M33 we could do extremely well by using a two-color diagram to separate the
foreground stars from the RSGs, as $B-V$ is sensitive to surface gravity while $V-R$ remains a temperature indicator [13-15].  Were it not for this, the foreground contamination would
be about 80\% [16].

Having successfully identified clean samples of these stars, how many did we actually find, and what did we learn by doing so?

\subsection{YSGs}
Samples of 176 and 317 YSGs have been identified in the SMC and LMC, respectively [10, 11], and samples of 120 and 121 YSGs have been found in M31 and M33, respectively [12, 13].
These samples are not necessarily complete spatially, but care was taken to make the samples complete in luminosities in each galaxy, extending down to 12 $M_\odot$ ($\log L/L_\odot \sim 4.5$).  This allowed us to compare the relative number of YSGs as a function of mass to those predicted by the single-star evolutionary models.  

We began by comparing the relative numbers of YSGs as a function of luminosity and mass in M31 [12]. The models predicted lifetimes during the YSG phase that were
actually longer for the higher mass stars, and even when the numbers were decreased by the expected correction for the initial mass function, there was a large discrepancy.
Of course, the evolutionary time scale for the YSG phase is quite short, of order a mere 10,000 years, and extremely sensitive to a number of factors.  We were curious if the
problem was related to the assumed mass-loss rates at the high metallicities found in M31, and so we repeated the test in the lower metallicity SMC [10] with very similar results.
By the time we made a comparison in the LMC [11] and in M33 [12], there was a new generation of evolutionary models available [6].  The agreement with the
predicted relative numbers were in spectacular agreement with these new models [11-12].   As usual in our collaborations
with the Geneva evolutionary group, the approach to fix problem is not to tweak parameters, but rather to make the physics less approximate. The improvements incorporated in the newer models included improved opacities (due to
revised compositions), updated reaction rates, a new prescription for mass-loss rates during the RSG phase, and new shear diffusion coefficients [6]. The better agreement with observations is likely due to a combination of these effects rather than a single one as argued by [13]. 

\subsection{RSGs}

\begin{figure}[!h]
\centering\includegraphics[width=2.5in]{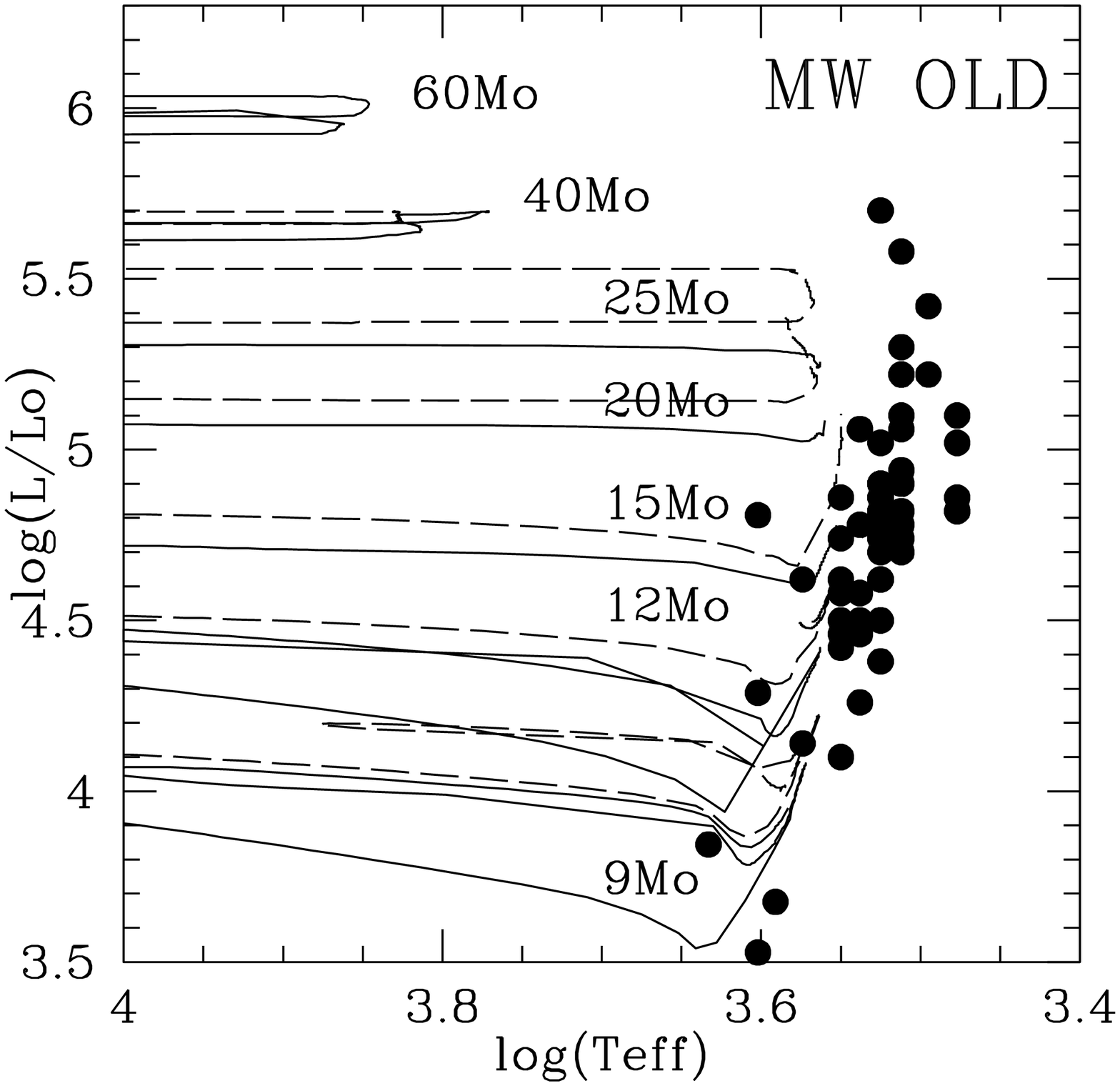}
\centering\includegraphics[width=2.5in]{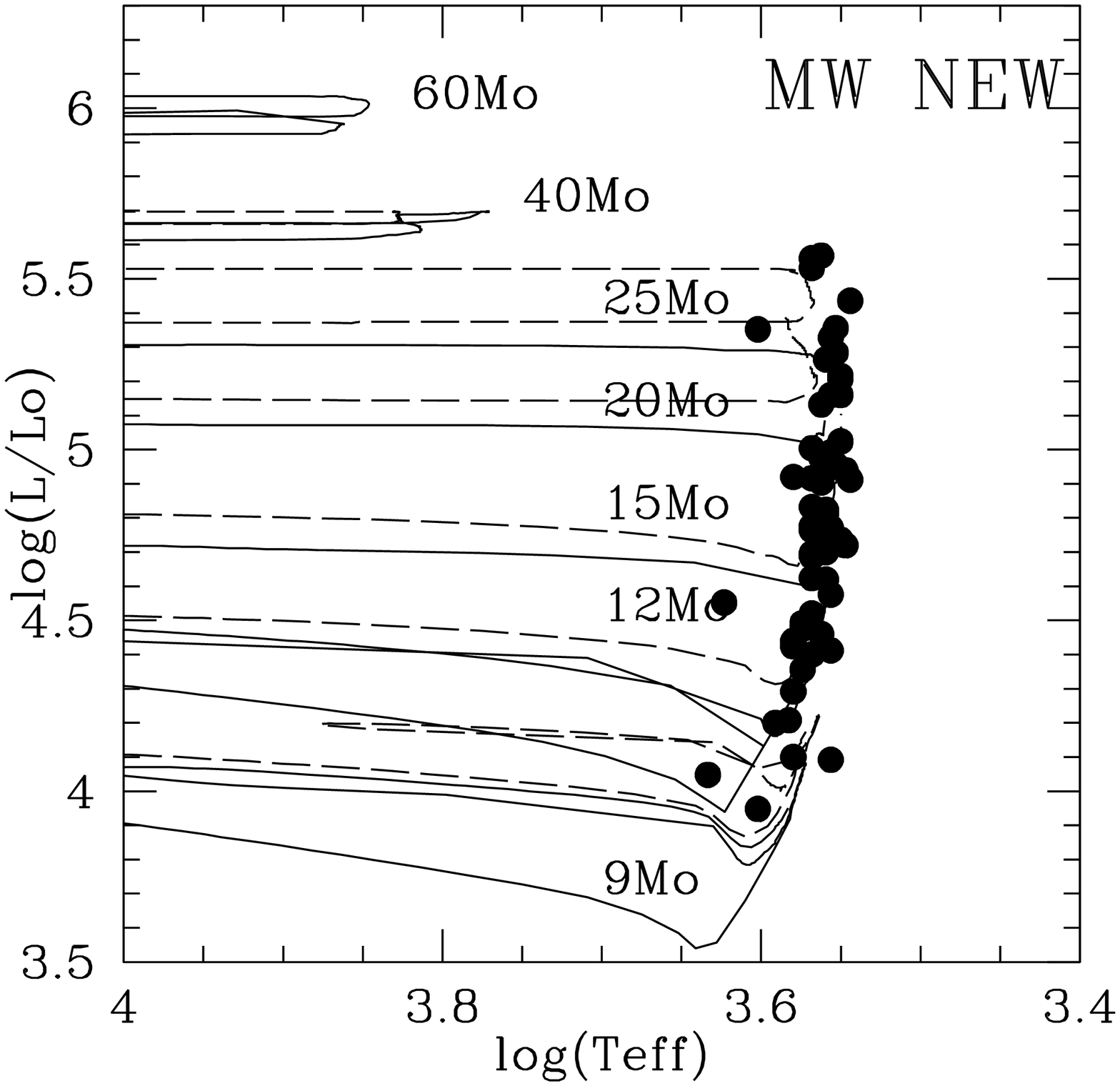}
\caption{The improvement in the effective temperature scales of RSGs by [18] resulted in much better agreement with
the evolutionary tracks.  From [19]; used
with permission.}
\label{fig_emily1}
\end{figure}

Red supergiants are in some ways an even more neglected evolutionary stage than the YSGs.  Probably this is because the physics of their atmospheres are even more complicated than those of WRs.  Fifteen years ago we had a serious problem but no one seemed to even be aware of it: the ``observed" locations of RSGs in the H-R diagram
were cooler and more luminous than the evolutionary tracks predicted [17].   Usually when there is a discrepancy between theory and observations, we observers naturally try to point our fingers at the theorists.
In this case, we were suspicious that the problem may lay with the interpretation of the observations: how were spectral types (or colors) converted to effective temperatures?  A new generation of the MARCS stellar atmosphere models that included sphericity, improved opacities, and low surface gravities were used by [18] to determine reddenings and effective temperatures for a large sample of Galactic RSGs whose distances were known. When the improved parameters were used to compare to the evolutionary tracks, there was a substantial improvement (!), as shown in Fig.~\ref{fig_emily1}.
This model fitting relies upon fitting the TiO bands in the spectrum, the same lines used for the spectral types.  These lines become stronger with decreasing effective temperatures, as the spectral types becomes later. 

 RSGs cluster along a
vertical line (as shown in Fig.~\ref{fig_emily1}) known as the Hayashi limit [20].  Cooler than that, a star will no longer be in hydrostatic equilibrium.  The location of the Hayashi limit shifts to warmer temperatures as the metallicity decreases, and this is consistent with the fact that the average spectral type shifts to earlier types in lower metallicity systems,
as first explained by [21].  Recently the effective temperature scale of [18] and subsequent work [22] has been called into question by [23].  Their results suggest that all RSGs have essentially the same temperature regardless of spectral
type.  It is not clear how readily this can be reconciled with the shifting of the Hayashi limit as a function of metallicity (see, e.g., [22]).

The lifetimes of RSGs are much longer than that of YSGs, as basically the entire He-burning phase of stars 10-30$M_\odot$ are spent as RSGs.  There are $\sim$ 500 RSGs known in M31 [15, 16] and the LMC [11, 24], and about 150
known in NGC 6822 [25], the SMC [24], and M33 [13].  About a dozen are known in WLM [25].  Work is in progress by two of the present authors to identify and characterize these stars in Sextans A and B.

\section{Wolf-Rayet Stars}

The highest mass stars ($>30M_\odot$) spend their He-burning lives as WRs. In the single-star evolutionary model, mass-loss due to strong stellar winds during the main-sequence stage (plus additional mass loss during an LBV stage?) strip off the H-rich outer layers, revealing the H-burning products He and N.  Such a star is identified as a WN-type WR.  If enough additional mass loss occurs, then the He-burning products C and O are revealed, resulting in a WC-type WR. In the binary evolution scenario, this mass loss is accomplished primarily by Roche-lobe overflow.   

One of the burning questions in the study of WRs is the relative importance of these two mechanisms. It has been known since the pioneering work of [26] that the close binary frequency of nearby O-type stars is
about 35\%. An identical result was found for massive stars in 30 Dor region by [27]  and in Galactic OB associations [28].
The extrapolation of this result to the statement that ``binary interaction dominates the evolution of massive
stars" requires extrapolation to extremely long-period systems. While long-period (years) systems may 
interact when one component undergoes a RSG phase and becomes a bloated behemoth, the problem with
the conclusion that ``all" WRs must come from binary evolution is that few WR progenitors are expected to ever go through a RSG phase: the evolutionary tracks at higher masses ($>30M_\odot$) turn back to higher temperatures long before reaching a RSG phase. In other words, the WR progenitors never get large enough to interact with distant companions.  That is not to say that binary evolution may not provide an important channel to the formation of WRs.  But, the importance of binary evolution may be overstated these days.

The number and types of WRs provide important observational constraints on this question, as long as we can  identify complete (bias-free) samples.  Here the difficulty is that the strongest optical emission line in 
WCs (C III $\lambda 4650$) is
about 10$\times$ stronger than the strongest emission line in WNs (He II $\lambda 4686$), making WCs much easier to find (see, e.g., [29, 30]).  Historically, most WRs in the Milky Way and Magellanic Clouds had been found by objective prism surveys, random discovery by spectroscopy, or as part of directed searches using interference filter imaging. (See [19] for a more detailed summary).  

Complete surveys have now been carried out for WRs in M33 and M31, identifying 206 and 154 WRs, respectively in these galaxies [7, 30, 31].  These studies have used narrow-band (50\AA) interference imaging through filters centered
on C III $\lambda 4650$, He II $\lambda 4686$, and neighboring continuum ($\lambda 4750$), and combines the use of photometry and image-subtraction techniques to identify viable candidates, which are then examined spectroscopically.

A similar survey is underway for WRs in the Magellanic Clouds [32, 33, 34].  So far an additional dozen WRs have been discovered in the LMC (bringing the total number of LMC WRs to 154),
with all of the new stars of WN type.  However, the most amazing result is that ten of
these are of a type never before seen, which we are calling WN3/O3.  These stars show the emission characteristic of a WN3 star and the absorption spectrum of an O3; modeling has shown that the emission and absorption arise in the same object.  These stars still show an appreciable amount of hydrogen but nitrogen abundances typical of normal WRs.
They are hotter and fainter visually than other WRs, but have similar bolometric luminosities [35].  Further work on these interesting objects is underway by co-author K. Neugent as part of her thesis.  This work had further demonstrated the successes and limitations of our earlier surveys [36].

\begin{figure}[!h]
\centering\includegraphics[width=2.5in]{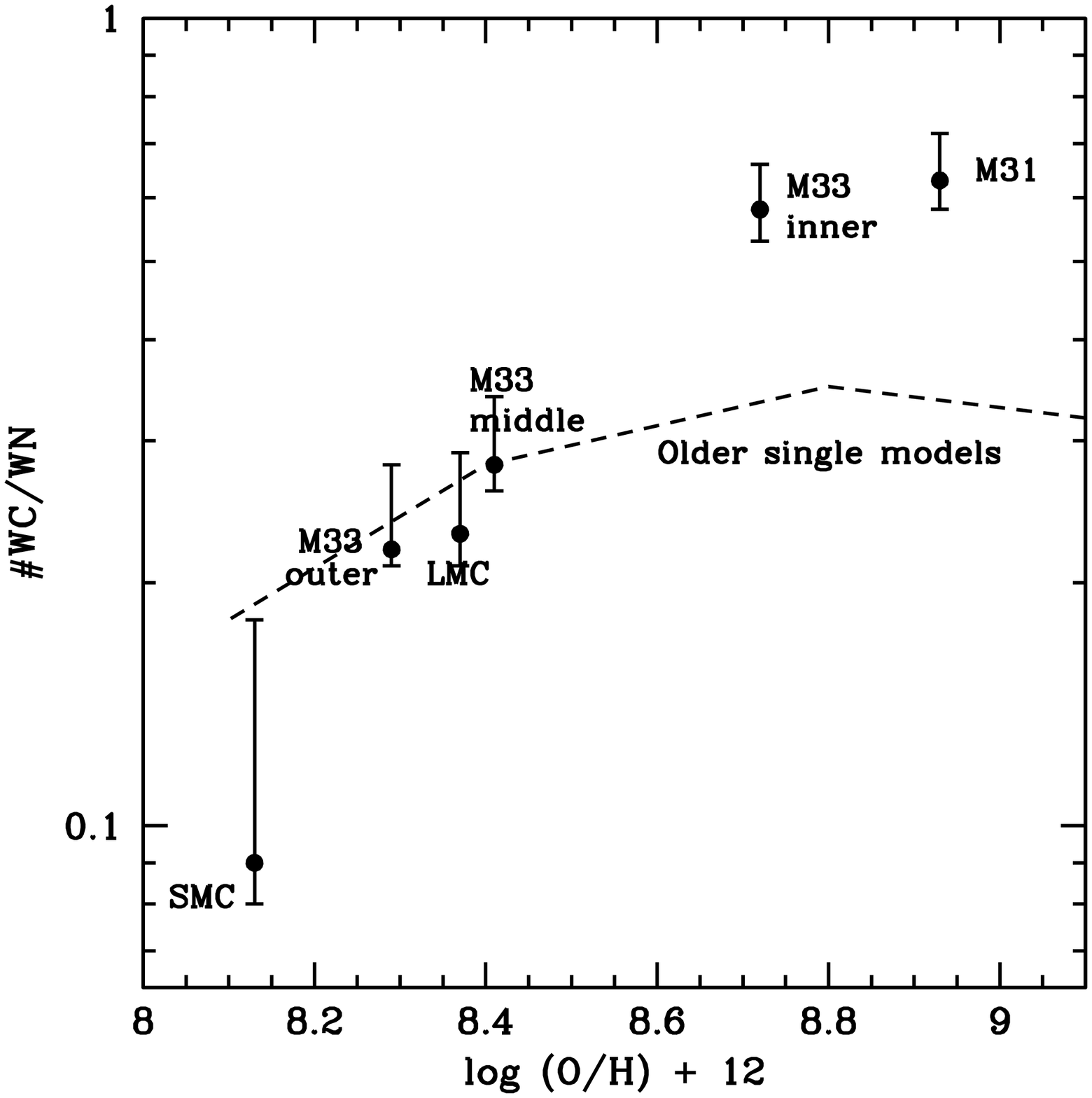}
\centering\includegraphics[width=2.5in]{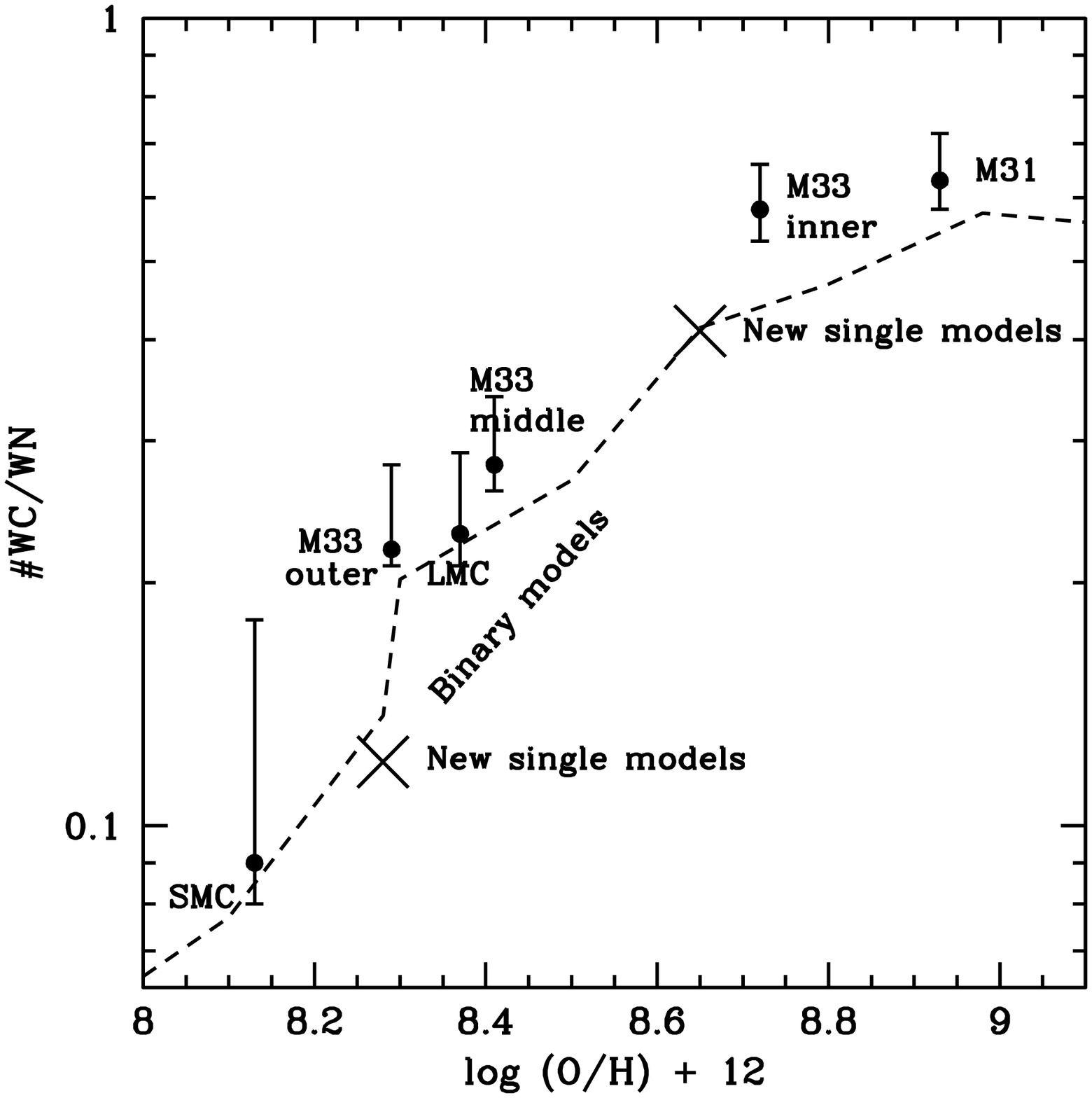}
\caption{The observed WC/WN ratios from [31] as a function of metallicity are compared with the predictions of
evolutionary models.  {\it Left:} the models shown are the older single-star Geneva models from [1].  {\it Right:} the line shows the predictions from the BPASS 2.0 models of J. J. Eldridge.  The two crosses are from the newer Geneva single-star models as given in [8] and [31].} 
\label{fig:WCWN}
\end{figure}

What else have we learned?  The most interesting result is shown in Fig.~\ref{fig:WCWN} (left) where we compare the
number ratio of WC and WN stars as a function of metallicity with those predicted by the single star models. 
We see here that the models do a good job at predicting this ratio at lower metallicities but not at higher. What then
could be the cause of the problem?

One possibility considered by [31] was that binary evolution might play a more important role at higher metallicities. What if the binary frequency of massive stars were higher at higher metallicity?  Thus, a study of the {\it relative} binary frequency of WR stars throughout M33 and M31 was carried out by [37], who found that no significant differences with metallicity. So, that does not appear to be the answer.

However, the BPASS 2.0 binary evolutionary models of J. J. Eldridge do a very good job of matching the behavior
of the WC/WN number ratio as a function of metallicity, as shown in Fig.~\ref{fig:WCWN} (right).  (We are
grateful to Dr. Eldridge for help in determining the WC/WN ratios from his models.) 
Does this then settle
the matter?  Unfortunately, no.  We include in this figure the predictions of the latest Geneva single-star models
[8, 31].  We see that those predictions agree almost perfectly with the binary ones!   It will be interesting to see if
this degeneracy persists as improved single-star models at higher metallicities become available, and when the effects of
stellar rotation are included in the BPASS models. 

\section{Conclusion}
Although the conference organizers may have intended for us to consider a different ``gap," here we have tried to summarize some of the observational gaps between massive star observations and massive star theory.  As both the observations and modeling improve, we expect (hope) that this gap will continue to narrow.  In the meanwhile, we leave you with the local admonishment to please ``mind the gap." 

\vskip6pt

\enlargethispage{20pt}

\ethics{}

\dataccess{}

\aucontribute{All three authors were heavily involved in the research discussed here.  PM drafted the manuscript, and all authors commented on, read, and approved the manuscript.}

\competing{The authors declare that they have no competing interests.}

\funding{PM's contributed have been support by Lowell Observatory and through the National Science Foundation through AST-1612874; KFN's contributions have been support in part from
program number GO-13780 provided by NASA through a grant from the Space Telescope Science Institute, which is operated by the Association of Universities for Research in Astronomy, Incorporated, under NASA contract NAS5-26555.}



\end{document}